\documentstyle[12pt]{article}

\def\Box{\hbox{$\sqcup$\kern-0.66em\lower0.03ex\hbox{$\sqcap$}}}

\begin{document}
\begin{titlepage}
\begin{flushright}
IFUP--TH 32/97
\end{flushright}
\vskip 1truecm
\begin{center}
\Large\bf
Group theoretical derivation of Liouville action for Regge surfaces
\footnote {This work is  supported in part
  by M.U.R.S.T.}.
\end{center}

\vskip 1truecm
\begin{center}
{ Pietro Menotti } \\
{\small\it Dipartimento di Fisica dell'Universit\`a, Pisa 56100,
Italy and}\\
{\small\it INFN, Sezione di Pisa}\\
\end{center}
\vskip .8truecm
\begin{center}
July 1997
\end{center}
\end{titlepage}

\begin{abstract}
We show that the structure of the Liouville action on a two
dimensional Regge surface of the topology of the sphere and of the
torus is
determined by the invariance under the transformations induced by the
conformal Killing vector fields and under modular transformations.
\end{abstract}
\section{Introduction}

The Liouville action plays a key role in two dimensional gravity. It
provides the quantum theory with a highly non trivial dynamics,  while
the corresponding classical theory is completely trivial.

On the continuum the Liouville action has been derived in
\cite{allc} and it is
a non perturbative result. On a two dimensional Regge surface, i.e. on
a surface which  is everywhere flat except at isolated points, the
exact expression of the Liouville action has been derived in
\cite{pmppp}. The
interest in considering Regge surfaces is that they provide the
reduction of the functional integral to a finite number of degrees of
freedom with a simple geometrical meaning \cite{pmpppsanta}.
The derivation \cite{pmppp}, employs the
$Z$-function regularization of functional determinants on singular
Riemann surfaces and the
Riemann-Roch theorem to fix the correct self-adjoint extension of the
conformal Lichnerowich-De Rahm operator.

One wonders whether a simpler argument can explain the structure of
the action.  What we shall show here is that, under reasonable
assumptions
that we shall state in the following, the structure of
the Liouville action for a Regge surface is dictated by group theory
alone, i.e. by the invariance of
the action under the transformations induced by the conformal Killing
vector fields and under modular transformations.

\section{Structure of the discretized Liouville action}
The Liouville action in Polyakov covariant non local form is given by

\begin{equation}
\displaystyle
\label{contliouv}
S_l = \frac{26}{96\pi} \left\{\int  d^{2}\omega \, d^{2}\omega'
\;(\sqrt{g}  R)_{\omega }
\frac{1}{\Box}(\omega, \omega')(\sqrt{g} R)_{\omega'}
 - 2 (\log {A\over A_0})
\int \: d^{2} \omega\,\sqrt{g} R \right\}
\end{equation}
where in the case of spherical topology
\begin{equation}
\frac{1}{\Box} (\omega,
\omega')=\frac{1}{2\pi} \log | \omega - \omega' |
\end{equation}
and
$A$ is the area of the surface.
$A_0$ is the reference area i.e. the value of the area due to the
$\sigma_0$ in the following decomposition of the conformal factor
$\sigma$
\begin{equation}
\sigma(\omega) = \sigma_0(\omega)+\lambda_0, ~~\sigma_0 \approx
-2\log|\omega| ~~{\rm for~}\omega\rightarrow \infty.
\end{equation}
The last term in eq.(\ref{contliouv}) in important as it provides the
correct
transformation properties of the action under uniform rescaling
\cite{duffdowker}.

In \cite{pmppp} the Liouville action has been derived for a two
dimensional
Regge surface, i.e. a surface which is everywhere flat except at
isolated points where  conical
singularities are present;
the result for spherical topology was
\begin{equation}
  \label{discrliouv}
S_l=  \displaystyle {\frac{26}{12} \left\{ \sum_{i,j\neq i}
    \frac{(1-\alpha_{i})(1-\alpha_{j})}{\alpha_{i}} \log |w_{i} -
    w_{j}| + \lambda_{0} \sum_{i} (\alpha_{i} - \frac{1}{\alpha_{i}})
    - \sum_{i} F(\alpha_{i}) \right\}}
\end{equation}
where $\alpha_i$ are the apertures of the conical singularities
($\alpha=1$ is the plane), $\omega_i$ the locations of the
singularities and $\lambda_0$ the scale parameter occurring in the
Regge conformal factor (see eq.(\ref{conformalfactor})
below). $F(\alpha)$
is a smooth
function for which an integral representation can be given. A similar
formula can be derived for the torus topology (see \cite{pmppp} and
eq.(\ref{torusliouv}) below).
The idea of using the complex plane with conical singularities to
describe a Regge  geometry is found in \cite{foer}\cite{aur}, while
the
derivation of eq.(\ref{discrliouv}) was given in \cite{pmppp} by
extending a
technique developed
in \cite{aur}. We recall that the description by means of the complex
plane is completely
equivalent to the one in terms of the bone lengths, but the former in
two dimensions looks more powerful.

We shall start from the topology of the sphere. As usual we shall
describe the surface with the topology of a sphere by a single chart
given by the complex plane completed by the point
at infinity (Riemann sphere) and the conformal factor describing the
Regge geometry is given by \cite{foer},\cite{aur},\cite{pmppp}
\begin{equation}
\label{conformalfactor}
e^{2\sigma} = e^{2\lambda_{0}} \prod_{i=1}^{N} | \omega - \omega_{i}
|^{2(\alpha_{i} -1)}, \quad \sigma \equiv \sigma(\omega ; \lambda_{0},
\omega_{i}, \alpha_{i} )=
\lambda_0+\sum_i(\alpha_i-1)\log|\omega-\omega_i|
\end{equation}
with the restriction $\sum_{i=1}^{N} (1 - \alpha_{i}) = 2$, i.e. the
sum of the deficits must be equal to the Euler characteristic.
Due to the presence of six conformal Killing vector fields for the
sphere
topology such a conformal factor is unique \cite{ginsparg} up to an
$SL(2,C)$
transformation.
Under
\begin{equation}
  \omega' = \frac{\omega a + b}{\omega c + d} \; , \quad  \omega =
  \frac{\omega'd - b}{- \omega' c + a} \; , \quad  ad - bc = 1
  \label{sl1}
\end{equation}
the new conformal factor is given by
\begin{equation}
\sigma'(\omega'; \lambda_{0}, \omega_{i}, \alpha_{i}) =
\sigma(\omega';  \lambda_{0}', \omega_{i}', \alpha_{i}')
\end{equation}
with
\begin{equation}
\label{conftras}
\lambda_{0}' = \lambda_{0} + \sum_{i=1}^{N} (\alpha_{i} -1 ) \log
|\omega_{i} c + d|, \quad
\displaystyle \omega_{i}' = \frac{a\omega_{i} + b}{c\omega_{i} + d },
\qquad \alpha_{i}' = \alpha_{i} \, .
\end{equation}
The area $A$
\begin{equation}
A=e^{2\lambda_0}\int d^2\omega |\omega-\omega_i|^{2(\alpha_i-1)}
\end{equation}
being a geometric invariant is left unchanged.

We now must construct a discrete transcription of the non local action
(\ref{contliouv}) in terms of the $3N$ parameters $\omega_{1x},
\omega_{1y},
\dots\omega_{Nx},\omega_{Ny},\alpha_1\dots\alpha_{N-1},\lambda_0$
characterizing the Regge conformal factor. The Green function on the
plane is
\begin{equation}
\frac{1}{2\pi}\log|\omega_i-\omega_j|
\end{equation}
and the curvature is concentrated at the points $\omega_i$ and
characterized by $\alpha_i$. As the integrated curvature at the
singularity $\omega_i$ is given by $4\pi (1-\alpha_i)$ it would be
natural to consider for the action the combination
$(1-\alpha_i)(1-\alpha_j)\log|\omega_i-\omega_j|$. However, while
providing the right behavior
for small angular deficit, i.e. small $1-\alpha_i$, it is easily
checked (and it will be proved in the following) that the ensuing
structure is not correct as it is not invariant under $SL(2,C)$.
Thus we must consider the more general structure
\begin{equation}
\sum_{i,j} K_{ij}[\alpha] \log|\omega_i-\omega_j| + B(\lambda_0,\alpha)
\end{equation}
where $K_{ij}[\alpha]$ has the following properties
\begin{equation}
K_{ij}[\alpha]= K_{ji}[\alpha],~~~~ K_{ii}[\alpha]=0.
\end{equation}
In fact $K$ has to be a symmetric function of $i,j$ which depends only
on the (singular) curvatures.
$B$ is a function
to be determined
of the $\alpha$'s and the scale parameter $\lambda_0$.
Under the $SL(2,C)$ transformation (\ref{conftras}) we have
\begin{equation}
|\omega'_i-\omega'_j| = \frac{|\omega_i-\omega_j|}{|c\omega_i+d|
|c\omega_j+d|}.
\end{equation}
Invariance of the action under such a transformation gives
us
\begin{equation}
-2 \sum_{i,j} K_{ij}[\alpha] \log|c \omega_i+d | +
B(\lambda'_0,\alpha)=B(\lambda_0,\alpha).
\end{equation}
Let us consider first a dilatation i.e. a transformation with
$a=1/d=\Lambda,~b=c=0$. We have
\begin{equation}
2 \sum_{i,j\neq i} K_{ij}[\alpha] \log \Lambda +
B(\lambda_0 +2 \log \Lambda ,\alpha)=B(\lambda_0,\alpha).
\end{equation}
The above relation tells us that $B$ is a linear function of
$\lambda_0$ i.e.
\begin{equation}
B(\lambda_0,\alpha)= -\lambda_0 \sum_{i,j}K_{ij}[\alpha] + F[\alpha].
\end{equation}
Thus we have reached the structure
\begin{equation}
\label{structure}
\sum_{i,j} K_{ij}[\alpha] \log|\omega_i-\omega_j| -\lambda_0
\sum_{i,j}K_{ij}[\alpha] +F[\alpha]
\end{equation}
We impose now the invariance of the above action under the general
transformation (\ref{sl1})
\begin{equation}
0 = 2 \sum_{i,j} K_{ij}[\alpha] \log|c\omega_i+d|+ \sum_i (\alpha_i-1)
\log|c\omega_i+d| ~~\sum_{m,n}K_{mn}[\alpha].
\end{equation}
It is useful to define $K_{ij}[\alpha]=
(1-\alpha_i)(1-\alpha_j)H_{ij}[\alpha]$ and, as the $\omega_i$ are
independent variables we obtain
\begin{equation}
\label{H}
2\sum_{j} H_{ij}[\alpha] (1-\alpha_j)=
\sum_{m,n}(1-\alpha_m)(1-\alpha_n)H_{mn}[\alpha].
\end{equation}
Notice that the second member does not depend on $i$.
We now specify that $K_{ij}[\alpha]$ depends on $\alpha_i$ and
$\alpha_j$ i.e. only on the (singular) curvatures at $\omega_i$ and
$\omega_j$. This means that $H_{ij}[\alpha]=h(\alpha_i,\alpha_j)$.
Then eq.(\ref{H}) becomes
\begin{equation}
\label{h}
2\sum_{j} h(\alpha_i,\alpha_j) (1-\alpha_j)=
\sum_{m,n}(1-\alpha_m)(1-\alpha_n)h(\alpha_m,\alpha_n).
\end{equation}
The above equation must hold for any choice of $\alpha_i >0$
respecting the Euler relation $\sum_i(1-\alpha_i)=2$ and for
any number $N$ of vertices. Given
$\alpha_1$  and $\alpha_2$ at will we shall choose all the other
$\alpha$'s equal and thus given by
\begin{equation}
1-\bar\alpha=\frac{\alpha_1+\alpha_2}{N-2}.
\end{equation}
Substituting into (\ref{h}) we have for $i=1$
$$
\alpha_1 (1-\alpha_2) h(\alpha_1,\alpha_2) + \alpha_1
(\alpha_1+\alpha_2) h(\alpha_1,\bar\alpha)=
$$
\begin{equation}
\label{hN}
(1-\alpha_2)(\alpha_1 + \alpha_2) h(\alpha_2,\bar \alpha)+
\frac{(N-3)}{2(N-2)}(\alpha_1+\alpha_2)^2 h(\bar \alpha, \bar\alpha).
\end{equation}
The above equation has to hold for any $N$ and thus also in the limit
$N \rightarrow \infty$ i.e.
\begin{equation}
\label{hinfty}
\alpha_1 (1-\alpha_2) h(\alpha_1,\alpha_2) + \alpha_1
(\alpha_1+\alpha_2)h(\alpha_1,1) =
(1-\alpha_2)  (\alpha_1+\alpha_2) h(\alpha_2,1) +
\frac{(\alpha_1+\alpha_2)^2}{2}h(1,1).
\end{equation}
The equation is homogeneous in $h$ and so we shall not be able,
using simply group theory to fix the
overall proportionality constant. We shall normalize $h$ to
$h(1,1)=2$. A solution to eq.(\ref{hinfty}) (and also to eq.(\ref{hN}))
is given by
\begin{equation}
h(\alpha_1, \alpha_2) = \frac{1}{\alpha_1}+\frac{1}{\alpha_2}.
\end{equation}
We shall put
\begin{equation}
h(\alpha_1, \alpha_2) = \frac{1}{\alpha_1}+\frac{1}{\alpha_2}+
r(\alpha_1,\alpha_2)
\end{equation}
with obviously $r(1,1)=0$, and our aim is to prove that
$r(\alpha_1,\alpha_2)$ is identically zero.
The equation for $r(\alpha_1,\alpha_2)$ now is
\begin{equation}
\label{req}
\alpha_1 (1-\alpha_2) r(\alpha_1,\alpha_2) + \alpha_1
(\alpha_1+\alpha_2)r(\alpha_1,1) =
(1-\alpha_2)  (\alpha_1+\alpha_2) r(\alpha_2,1)
\end{equation}
Putting in eq.(\ref{req}) $\alpha_2=1$ we have
\begin{equation}
\alpha_1 (\alpha_1+1)r(\alpha_1,1)=0
\end{equation}
from which $r(\alpha_1,1)=0$ and substituting into eq.(\ref{req}) we
get finally $r(\alpha_1,\alpha_2)=0$.

With regard to the function $F[\alpha]$ appearing in
eq.(\ref{structure}), being the
$\alpha$'s invariants under $SL(2,C)$ transformations, nothing can
be said from group theory alone; the only thing we notice here is that
as it happens for the exact discretized action (\ref{discrliouv}) for
small angular
deficits, i.e. near the continuum limit it becomes
\begin{equation}
F[\alpha]\approx F[1] + \frac{\partial F}{\partial \alpha_1}[1]
\sum_{i}(\alpha_i-1) = F[1]-\chi \frac{\partial F}{\partial
\alpha_1}[1]
\end{equation}
i.e. a topological invariant.

Once the non perturbative structure in known, the overall
coefficient $26/24$ can be borrowed from perturbation theory as was
originally done by Polyakov \cite{allc}. We recall however that the
procedures \cite{allc}
and \cite{pmppp} provide also the correct normalization.

In conclusion for the sphere topology we have reproduced
eq.(\ref{discrliouv}) except for the last term which in both formulae
becomes a topological invariant in the continuum limit.

For the torus the invariance group is confined to the modular group times
the translations.

The conformal factor describing a Regge surface with the topology of
the torus is given by \cite{pmppp}
\begin{equation}
\label{sigmatoro}
\sigma (\omega) = \lambda_{0} + 2\pi \sum_{i=1}^{N} (\alpha_{i} -1)
G(\omega-\omega_i|\tau)
\end{equation}
where $\omega = x+\tau y$,  being the fundamental region given by the
square $ 0\leq x <1,~ 0\leq y <1$  and $G(\omega-\omega'|\tau)$ is the
Green
function on the torus \cite{minzub}
\begin{eqnarray}
& \Box G(\omega - \omega' |\tau) = \delta^{2}(\omega - \omega') -
\frac{1}{\tau_{2}} & \\
& G(\omega - \omega' | \tau)  = \frac{1}{2\pi} \log \left|
    \frac{\vartheta_{1} (\omega -
\omega'| \tau )}{\eta(\tau)} \right| - \frac{(\omega_{y} -
\omega_{y}')^{2}}{2 \tau_{2}}  &.
\end{eqnarray}
$\vartheta_{1}(\omega | \tau)$ is the Jacobi $\vartheta$--function
and
\begin{equation}
\displaystyle
\eta(\tau)=  e^{\frac{i\pi\tau}{12}} \prod_{n=1}^{\infty} [ 1 -
e^{2in\pi\tau} ] .
\end{equation}
The discrete structure inherited from (\ref{contliouv}) is
\begin{equation}
\label{discrtorus}
2\pi \sum_{i,j} K_{ij}[\alpha,\tau] G(\omega_i-\omega_j|\tau) +
B(\lambda_0,\alpha,\tau).
\end{equation}
We recall that $G(\omega_1 - \omega_2 | \tau)$ is modular invariant
i.e. under the modular transformation
\begin{equation}
\label{transmod}
\tau \longrightarrow \tau' = \frac{\tau a + b}{\tau c + d }
\end{equation}
with $(a,b,c,d) \in {\bf Z}$ and $ad -bc =1$ and
\begin{equation}
\label{modtrans}
\omega' = \frac{\omega}{\tau c + d}
\end{equation}
we have
\begin{equation}
G(\omega_1 - \omega_2 | \tau) = G(\omega_1' - \omega_2' | \tau').
\end{equation}
We shall assume now that at short distances the divergence
behavior of the action should be independent of the topology of the
manifold  i.e. for $\omega_1-\omega_2\rightarrow 0$ the discrete
Liouville action should diverge like on the sphere which, as
\begin{equation}
2\pi G(\omega_1 - \omega_2 | \tau) \approx \log|\omega_1 - \omega_2|
\end{equation}
imposes again
\begin{equation}
K_{12}[\alpha] = (1-\alpha_1)(1-\alpha_2) \left (
\frac{1}{\alpha_1}+\frac{1}{\alpha_2}\right).
\end{equation}
We must now determine the function $B(\lambda_0,\alpha,\tau)$.
Under the modular transformation (\ref{transmod}) and (\ref{modtrans})
$\lambda_0$ goes over to
\begin{equation}
\lambda_{0}' = \lambda_{0} + \log |\tau c + d | \, .
\end{equation}
Due to the modular invariance of the Green function we must have
\begin{equation}
B(\lambda_0,\alpha,\tau)=B(\lambda'_0,\alpha,\tau')=B(\lambda_0+ \log
|\tau c + d |,\alpha,\frac{\tau a + b}{\tau c + d }).
\end{equation}
If we write, without any loss of generality
\begin{equation}
\label{bcequ}
B(\lambda_0,\alpha,\tau)=C(\lambda_0-\log|2\pi
\eta^2(\tau)|,\alpha,\tau)
\end{equation}
recalling that $\eta^{24}(\tau)$ is a modular cusp form of
weight $12$ i.e.
\begin{equation}
\eta^{24}\left(\frac{\tau a + b}{\tau c + d }\right)=
(c\tau+d)^{12} \eta^{24}(\tau)
\end{equation}
we have that $C$ is invariant under modular transformations i.e.
\begin{equation}
C(x,\alpha,\tau)=C(x,\alpha,\frac{\tau a + b}{\tau c +
d }).
\end{equation}
In order to specialize further the function $C$ let us consider the
case when all $\omega_i$ are all close together. In this case the
geometry is
given by a local deformation with zero integrated curvature of a flat
torus of dimensions much larger than the size of the deformation.
In the described situation and $\omega$ near the $\omega_i$'s the
conformal factor (\ref{sigmatoro}) becomes
\begin{equation}
\label{conffact1}
\sigma(\omega) \approx \lambda_0 +\sum_i(\alpha_i-1)
\log|\omega-\omega_i|
\end{equation}
and as it is easily checked the conformal factors (\ref{conffact1})
and the following (\ref{conffact2})
\begin{equation}
\label{conffact2}
\sigma'(\omega') \approx \lambda_0 -\log k +\sum_i(\alpha_i-1)
\log|\omega'- k \omega_i|
\end{equation}
describe the same geometry.
Imposing in such a limit the invariance of the action
(\ref{discrtorus}) under
$\omega_i\rightarrow k \omega_i\rightarrow$ and $\lambda_0 \rightarrow
\lambda_0 - \log k$ we obtain, taking into account that for the torus
$\sum_i(1-\alpha_i)=0$
\begin{equation}
B=2 (\lambda_{0} - \log | 2\pi \eta^{2} | ) \sum_{i} (\alpha_{i} -
    \frac{1}{\alpha_{i}}) + 2 F[\alpha,\tau]
\end{equation}
with $F[\alpha,\tau]$ invariant under modular transformations.
Obviously any $F[\alpha,\tau]$, even discontinuous in $\tau$, provided
it is invariant under (\ref{transmod}) will satisfy our invariance
requirements. It is however of interest to see what happens if for
$e^{S_l}$ we remain in the domain of modular functions. Any modular
invariant function can be written as a rational function of the
modular invariant $j(\tau)$ \cite{serre}. Moreover $j(\tau)$
has a simple pole at infinity and
provides a bijection between $H/G$ onto the complex plane $C$, where
$H/G$ the quotient of the upper complex plane with the modular
group \cite{serre}.
Thus if we want to avoid singularities in $\tau$ we must choose $F$
independent of $\tau$.
With such a choice we have for
the torus
\begin{equation}
\label{torusliouv}
S_{l} =  \frac{26}{12}\left[\sum_{i,j\neq i} \frac{(1 - \alpha_{i})(1
    -\alpha_{j})}{\alpha_{i}} 2\pi G(\omega_i-\omega_j|\tau)
 + (\lambda_{0} - \log | 2\pi \eta^{2} | ) \sum_{i} (\alpha_{i} -
    \frac{1}{\alpha_{i}}) + F[\alpha]\right] \, .
\end{equation}
where again the normalization has been borrowed from perturbation
theory. Eq.(\ref{torusliouv}) is identical to the one derived in
\cite{pmppp} except for the last term $F[\alpha]$ which here is left
generic while in \cite{pmppp} is completely determined. However both
terms in the continuum limit go over to a topological invariant.
\section{Conclusions}

The complete determination of the Liouville action for a Regge surface
in two
dimensions has been given in \cite{pmppp} both for the sphere topology
and the
torus  topology and it was shown that, as expected,  the obtained
actions are
invariant under the continuous transformations induced by the
conformal
Killing vector fields and under the modular group. Here the question
has been addressed of how far such symmetries restrict the translation
of
the continuum Polyakov covariant non local expression of the
Liouville action to the discrete case.

Is is found that, under reasonable assumptions, the action for the
sphere and torus topology are completely fixed except for a function
of the deficit angles which in the continuum goes over to a
topological invariant and the overall normalization of the action
which has to be borrowed e.g. from perturbation theory.

Thus even if the present derivation does not replace the full one
given in
\cite{pmppp}, it provides a simpler treatment and shows that the
invariance
principles are practically sufficient
to fix the form of the action.

\bibliographystyle{plain}

\end{document}